\documentclass[11pt]{article}

\usepackage[utf8]{inputenc}
\usepackage[T1]{fontenc}
\usepackage[english]{babel}

\usepackage[a4paper,top=2cm,bottom=2cm,left=2.2cm,right=2.2cm,marginparwidth=1.75cm]{geometry}

\usepackage{setspace}
\usepackage{graphicx}
\usepackage{amsmath,amssymb}

\usepackage[numbers]{natbib}
\usepackage[colorlinks=true, allcolors=blue]{hyperref}

\usepackage{titlesec}
\usepackage{caption}
\usepackage{multicol}
\usepackage{ragged2e}
\usepackage[dvipsnames]{xcolor}
\usepackage[most]{tcolorbox}

\usepackage{authblk}

\makeatletter

\renewcommand\AB@affilsepx{\protect\\[0.2em]}

\renewcommand{\maketitle}{%
  \begin{center}
    {\LARGE\@title\par}
    \vskip 1em
    {\large
      \setlength{\parskip}{0.3em}%
      \setlength{\parindent}{0pt}%
      \@author\par}%
    \vskip 1em
    \vskip 1em
  \end{center}
}

\makeatother

\title{Why the Northern Hemisphere Needs a 30-40 m Telescope and the Science at Stake: Time-domain astronomy}

\author[1,2,3]{F.~Coti Zelati}
\author[4]{P.~G.~Jonker}
\author[1,2]{C.~P.~Guti\'errez}
\author[5]{S.~Mattila}
\author[6]{D.~Pollacco}
\author[1,2]{N.~Rea}
\author[1,2]{P.~Charalampopoulos}
\author[7,8]{M.~A.~P.~Torres}
\author[7,8]{T.~Muñoz~Darias}
\author[3]{M.~C.~Baglio}
\author[1,2]{L.~Galbany}
\author[7,8]{E.~Villaver}

\affil[1]{Institute of Space Sciences (ICE, CSIC), Campus UAB, Carrer de Can Magrans s/n, Barcelona, E-08193, Spain \href{mailto:cotizelati@ice.csic.es}{\texttt{cotizelati@ice.csic.es}}}
\affil[2]{Institut d’Estudis Espacials de Catalunya (IEEC), Barcelona, E-08034, Spain}
\affil[3]{INAF--Osservatorio Astronomico di Brera, Via Bianchi 46, I-23807 Merate (LC), Italy}

\affil[4]{Department of Astrophysics/IMAPP, Radboud University, 6525 AJ Nijmegen, The Netherlands}

\affil[5]{Department of Physics \& Astronomy, University of Turku, Finland}

\affil[6]{Department of Physics, University of Warwick, Gibbet Hill Road, Coventry
CV4 7AL, UK}

\affil[7]{Instituto de Astrofísica de Canarias, E-38205, La Laguna, Tenerife, Spain}
\affil[8]{Universidad de La Laguna, Departamento de Astrofísica, E-38206, La Laguna, Tenerife, Spain}

\begin{document}
\maketitle
\newpage

\begin{tcolorbox}[colback=RoyalBlue!5!white,colframe=black!75!black, width=\textwidth]
\justifying
{\noindent We outline the science case for a 30–40\,m optical/IR telescope in the Northern Hemisphere, optimised for transformative time-domain astronomy in the 2040s. Upcoming multi-wavelength and multi-messenger facilities will reveal fast, faint, rapidly evolving Northern transients whose earliest phases carry decisive diagnostics. A Northern ELT with rapid response, broad wavelength coverage, high time resolution, polarimetry, and diffraction-limited imaging is essential to capture these phases and secure deep spectroscopy and photometry as transients fade. These capabilities will enable recovery of key physical information and detailed characterisation of transient environments, while also enabling unprecedented studies of accretion phenomena at all scales. La Palma uniquely combines atmospheric stability, complementary longitude to ESO’s ELT, protected dark skies, and robust infrastructure to host this facility.}
\end{tcolorbox}

\vspace{-0.5cm}
\section{Introduction and Motivation}
\vspace{-0.32cm}
Time-domain astronomy in the 2040s will be driven by continuous alert streams from wide-field, multiwavelength surveys and next-generation gravitational-wave (GW; e.g., Einstein Telescope, Cosmic Explorer, LISA) and neutrino detectors. 
For many faint explosive transients, key signatures are confined to the earliest phases and are rapidly lost as the ejecta expand and cool, making later observations far less constraining.
Extracting physical insight from events will thus increasingly rely on optical/IR follow-up within hours of the trigger, when key diagnostic information remains accessible. At the same time, extreme sensitivity will be crucial to probe fainter late-time components and to characterise host environments.
Yet, both ELT-class observatories under construction (ESO's 39\,m ELT and the Giant Magellan Telescope) are in the Southern Hemisphere (and at nearly the same longitude), from which a substantial fraction of the extragalactic transient sky is obscured by the Galactic Bulge and Centre. Moreover, without a Northern ELT, transients for example in key nearby galaxies (e.g. M31, M33, M81, M82, M101, Arp 299) will be inaccessible or will lack ELT-class sensitivity during their most diagnostic phases. This hemispheric imbalance is further amplified by the fact that many upcoming facilities (e.g. CHORD, ngVLA, LOFAR2.0, CTAO-North) will operate in the North, and IceCube-Gen2 will be most sensitive to up-going neutrinos from the Northern sky.

A Northern Hemisphere ELT with rapid-response capability, broad wavelength coverage, high-time-resolution instrumentation, polarimetric functionality, and advanced adaptive optics (AO) would be transformative for transient studies, while providing capabilities complementary to ESO’s ELT. 
Located on La Palma, such a facility would also offer critical longitudinal complementarity to Chile: it would enable earlier follow-up of fast transients that first reach optimal visibility from La Palma, and extended temporal coverage of rapidly evolving events visible from both hemispheres through coordinated observations as targets transition between the two sites’ night-time windows.

The transient classes outlined below are compelling in their own right, but also shape their environments and provide insight into black-hole (BH) formation and growth. Time-domain astronomy thus informs galaxy evolution, chemical enrichment, and even our understanding of the Epoch of Reionisation, making it broadly relevant to the astronomical community.

\vspace{-0.32cm}
\section{The Science Challenge}
\vspace{-0.32cm}
\noindent \textbf{Extragalactic transients} [\emph{What governs early optical/IR emission in fast transients?
Which progenitors and channels produce each class, and how do environment and geometry shape observables?
Do fast radio bursts have prompt optical/IR counterparts, and what powers them?}]

\noindent \underline{Gamma-ray bursts (GRBs)} provide direct insight into relativistic jets, compact-object physics, the most powerful cosmic explosions, and their environments (e.g., \cite{Levan2026}). Their earliest optical/IR emission encodes diagnostics of jet breakout, photospheric components, energy dissipation, and magnetic-field geometry. These phases remain sparsely sampled, yet they are key for understanding the central engine.
Duration alone cannot distinguish collapsars from compact-object mergers: some long GRBs show kilonova signatures, while some short GRBs are from massive-star explosions. Discriminating these pathways requires rapid, deep spectroscopy and photometry within the first hours–days, when kilonova components peak and shift into the IR. A northern 30–40\,m telescope would detect these faint, fast-evolving signals out to hundreds Mpc, enabling secure classification and \emph{r}-process diagnostics via high-resolution IR spectroscopy.
GRBs will also be central to multi-messenger astronomy in the era of future GW detectors, which will discover neutron-star (NS) mergers at $z \gtrsim 2$–3, where kilonova emission will be too faint to detect even with an ELT-class facility. For most events, afterglows will thus be the primary means to determine redshift, environment, jet structure, and magnetic-field geometry.
A northern ELT with rapid response, NUV–IR coverage, and high-throughput spectroscopy would enable routine early follow-up, delivering early-phase spectra probing dissipation mechanisms and spectropolarimetry constraining jet magnetisation. This capability would make possible systematic GRB/kilonova studies and provide the Northern complement needed to map jet physics, progenitors, and heavy-element production across cosmic time.

\noindent \underline{Supernovae (SNe).} 
For core-collapse SNe (CCSNe), shock breakout (SBO) is the first electromagnetic signal, encoding information on the progenitor radius, envelope structure, and circumstellar material (CSM)\citep{Jerkstrand2026}. The SBO emission signatures evolve within hours and peak in the UV/blue (only rarely extending to X-rays), making them largely inaccessible to current facilities. Early-time ($<$1\,d) high-resolution spectroscopy is uniquely diagnostic: transient flash-ionisation lines trace recent mass loss and reveal the kinematics and geometry of the CSM, including binary interaction signatures, but fade rapidly. For thermonuclear SNe (SNe Ia), early spectroscopy can distinguish between single- and double-degenerate progenitor channels and reveal the origin of early flux excesses\citep{Ruiter2025}. For both CCSNe and SNe Ia, early-time polarimetry can constrain ejecta geometry and asymmetries, testing for CSM-induced departures from spherical symmetry. 
Observations at $>$300\,d, when SNe have faded $>$3\,mag, are currently limited to nearby objects. A 30–40\,m-class telescope would enable NUV/optical–IR spectroscopy at these epochs, allowing measurements of progenitor composition, explosion asymmetries, ejecta–CSM interaction, and dust formation. 
At high redshift, ELT-class IR sensitivity is crucial to use CCSNe to trace the cosmic star-formation history, use SNe Ia to test redshift-dependent evolution and associated cosmological systematics, and identify rare pair-instability SNe from extremely metal-poor (Pop III) stars.
In synergy with Northern surveys and next-generation high-energy monitors, a Northern ELT would transform SN studies by linking progenitor evolution, explosion physics, and CSM interaction to finally understand stellar evolution.

\noindent \underline{Fast X-ray transients (FXTs)} encompass a wide range of phenomena and formation channels (e.g., SN SBOs, binary NS mergers, softer analogues of long GRBs, white-dwarf (WD) tidal disruption events, jet-driven explosions from massive stars \citep{Jonker2013,Quirola-Vasquez2023,Sun2025}). Over the past $\approx$1.5\,yrs, Einstein Probe has transformed the field by dramatically increasing the discovery rate ($\gtrsim$100 FXTs reported).
This surge in detections and the still small number of securely classified events underscores the need for rapid optical/IR follow-up to access key diagnostics and distinguish among progenitor scenarios: spectroscopy pinpoints redshift and probes the circum-burst environment; multi-band photometry constrains emission mechanisms and colour evolution; and integral-field spectroscopy characterises host galaxies and local environments.
In the 2040s, X-ray facilities such as NewAthena and transient-oriented missions like THESEUS and AXIS will provide FXT discoveries for ELT characterisation.

\noindent \underline{Luminous fast blue optical transients (LFBOTs)} are rare, luminous, rapidly evolving events with hot, blue, nearly featureless early spectra \citep{Margutti2019}. Their bright X-ray and radio emission as well as late-time rapid variability require powerful central engines and dense circumstellar environments, yet their nature remains debated. Proposed channels range from massive-star core collapse with fallback accretion to magnetar-powered explosions, tidal disruptions by intermediate-mass BHs, and exotic compact-object mergers. Discriminating among these scenarios requires rapid, high-S/N spectroscopy to track continuum cooling, emerging features, and ejecta kinematics, together with fast photometry and sensitive polarimetry during the engine-powered phase. A Northern 30-40-m facility can provide hour-cadence spectroscopy for faint events and deep late-time spectra to probe environments and hosts -- crucial capabilities as many LFBOTs will be found by Northern surveys.

\noindent \underline{Tidal Disruption Events (TDEs)}.
The earliest phases of TDEs are key to understand stellar-debris circularisation, accretion-disc formation, and the launch of winds or relativistic jets \citep{Gezari2021}, yet remain largely unobserved. High–S/N spectroscopy during the rise and early peak is required to track photospheric evolution, characterise the ionising continuum, and diagnose fast-evolving line and absorption features signalling outflows. A Northern 30–40-m facility would deliver early, high-throughput spectroscopy of faint, rapidly evolving TDEs at declinations unreachable from southern ELTs, providing the first direct view of disc formation and outflow launching around quiescent massive BHs. Moreover, time-domain surveys will uncover TDEs at higher redshifts, where sources are fainter, offering a unique probe of early supermassive and intermediate-mass BH growth. Spectroscopy will be essential to classify sources and train machine-learning photometric classifiers.

\noindent \underline{Fast Radio Bursts (FRBs)}. Despite mounting evidence linking some FRBs to magnetars, no prompt optical/NIR counterpart has yet been detected, likely due to limited sensitivity: current limits reach only $\sim$17\,mag on ms timescales for extragalactic repeaters and are thus consistent with most emission models. Facilities such as CHORD, DSA-2000, and BURSTT will vastly expand FRB discovery rates in the Northern sky. A northern 30–40\,m telescope equipped with ultra-fast photometers would enable ms optical monitoring of repeating FRBs, directly testing whether coherent radio bursts are accompanied by faint optical flashes and placing stringent constraints on emission efficiencies and radiative mechanisms \citep{Petroff2022,Zhang2023}. Diffraction-limited imaging and integral field units (IFU) spectroscopy of host galaxies would link burst properties to their local environments.

%\noindent \textbf{Accretion and ejection in compact objects} 
\noindent \textbf{Accretion at all scales} [\emph{How is angular momentum transported through accretion flows, and how do inflows couple to winds, jets, and magnetic fields across the mass scale? What drives the variability linking accretion and outflows?}]
Accretion discs power compact objects from accreting WDs in cataclysmic variables to NSs and BHs in X-ray binaries, and up to supermassive BHs in active galactic nuclei (AGN)\citep{Frank2002}. Despite decades of progress, the mechanisms transporting angular momentum and energy in discs (magnetised turbulence, instabilities, winds, and magnetosphere–disc coupling) remain poorly constrained. The key physics is encoded in rapid variability and correlated continuum and line responses, yet many decisive observables (e.g. sub-second variability, evolving line profiles, rapid polarisation changes) remain beyond current sensitivity and cadence, especially for faint systems.
A Northern 30–40\,m telescope would enable high-throughput, high-cadence optical/IR spectroscopy, photometry, and polarimetry of high-declination targets across accretion state, magnetic field, and mass-transfer rate. In accreting WDs, time-resolved spectroscopy can localise where variability is generated and how it propagates through discs. In NS and BH binaries, the same capabilities would track state transitions, probe disc truncation and wind launching, and isolate synchrotron jet components to map disc–jet coupling. Extending these studies to AGN tests whether characteristic variability timescales scale predictably with mass. Synergy with future Northern radio arrays and high-energy missions would enable coordinated campaigns on high-declination targets, delivering a unified, multi-wavelength view of accretion and outflows from WDs to supermassive BHs, including access to particle acceleration mechanisms in quickly spinning accreting NSs.% \citep{Papitto2022}.

\vspace{-0.32cm}
\section{Capability Requirements for Solving the Challenge}
\vspace{-0.32cm}
$\bullet$ \textbf{Deep sensitivity and rapid response:}
A $\sim$30-40-m aperture enables minute-scale spectroscopy and photometry of transients at $m \approx 23$–25. Achieving this in the earliest and most diagnostic phases requires low overheads, rapid target acquisition, and fast intra-night instrument switching.

\noindent $\bullet$ \textbf{Broad wavelength coverage:}
$0.3$–$14\,\mu$m coverage is required to track SBO continua, high-ionisation emission features, dust-obscured transients, and dust (pre-existing or newly-formed) re-radiation, as well as to probe rest-frame optical/UV emission from high-redshift sources. 

\noindent $\bullet$ \textbf{High-throughput spectroscopy:} 
Multiple spectral-resolution modes (low/medium for classification; high resolution for CSM and narrow-line diagnostics) are necessary for rapid, information-rich characterisation of transients. 

\noindent $\bullet$ \textbf{High-time resolution capability:}
Ultra-fast photometers with kHz–100\,kHz sampling rates, sub-ms to $\mu$s time stamping, simultaneous multi-band capability, and low-dead-time spectroscopic modes are essential to capture rapid variability, search for FRB optical counterparts, and enable the detection and timing of optical millisecond pulsars.
%These capabilities, already mature on smaller telescopes, would be transformative with ELT-class grasp.

\noindent $\bullet$ \textbf{Precision polarimetry and spectropolarimetry:}
High-throughput, low-systematics polarimetric and spectropolarimetric modes with rapid modulation are needed to measure evolving magnetic-field geometry, shock asymmetries, and jet/wind contributions in fast-evolving sources.

\noindent $\bullet$ \textbf{Advanced AO:}
Next-generation AO delivering diffraction-limited imaging and AO-fed IFUs are essential to resolve transient environments (particularly those embedded within the nuclear regions of their host galaxies). AO performance at or beyond the ESO ELT level is needed to enable precise host subtraction, localisation, and studies of intrinsically faint or high-redshift transients.

\noindent $\bullet$ \textbf{Multi-messenger integration \& flexible operations:}
Seamless integration with alert networks, programmatic ToO triggering compatible with community brokers, and automated priority-based scheduling are critical for time-sensitive events. Quicklook pipelines must deliver rapid classification, redshifts, and data-quality metrics to optimise coordinated follow-up.

\vspace{-0.32cm}
\section{Summary}
\vspace{-0.32cm}
In the 2040s, progress on fast, faint, and quickly evolving transients will require ELT-class, minute-to-hour optical/IR follow-up in the Northern Hemisphere. A 30-40\,m facility with the capabilities outlined above is essential to access the earliest, most diagnostic phases of explosive, multi-messenger, and accreting events. In combination with ESO's southern ELT, such a facility would deliver near-continuous global access to the transient sky. La Palma uniquely offers the latitude, atmospheric stability, longitudinal complementarity, and protected dark skies required to realise this vision.

\makeatletter
\renewcommand{\bibsection}{\section*{\refname}\vspace{-0.32cm}}
\makeatother
\vspace{-0.32cm}
\begin{multicols}{2}
\renewcommand{\bibfont}{\footnotesize}  
\bibliographystyle{unsrtnat}    
\bibliography{sample}
\end{multicols}

\end{document}